\documentclass[useAMS,usenatbib]{aa}

\def\draftversion{1} 
\usepackage{amssymb,latexsym,graphicx,natbib,eufrak,times,amsmath,xspace,ifthen, xcolor}
\usepackage[normalem]{ulem}
\usepackage{orcidlink}

\usepackage{hyperref}
\definecolor{linkblue}{rgb}{0,0.4,0.6}
\hypersetup{colorlinks, linkcolor={linkblue}, citecolor={linkblue}, urlcolor={linkblue}}

\ifthenelse{ \draftversion > 0 }{
  \newcommand{\so}[1]{\color[rgb]{0.6,0,0.6}\sout{#1}\color{black}} 

} {
  
  \newcommand{\so}[1]{}
}

\ifthenelse{\equal{\draftversion}{1}}{
  \newcommand{\sep}[1]{\par\begin{color}[rgb]{0,0.4,0}\begin{center}\hrule\end{center}\end{color}\par} 
  \newcommand{\todo}[1]{\begin{color}{red}\ \ifthenelse{\equal{#1}{}} {$\bullet\bullet\bullet$} {$\bullet$\ #1 $\bullet$}\end{color}} 
  \newcommand{\idea}[1]{\begin{color}[rgb]{0,0.4,0}\textit{#1}\end{color}} 
  \newcommand{\sk}[1]{\begin{color}[rgb]{0.6,0,0.6}#1\end{color}} 
  \newcommand{\toc}{\par\begin{color}[rgb]{0.6,0,0.6}\begin{center}\hrule\vspace{0.5mm}\begingroup\small\let\cleardoublepage\relax\let\clearpage\relax\mytoc\endgroup\vspace{0.5mm}\hrule\end{center}\end{color}\par} 

  }{
  \newsavebox{\trashcan}

  \newcommand{\sep}[1]{}
  \newcommand{\todo}[1]{}
  \newcommand{\idea}[1]{}
  \newcommand{\sk}[1]{}
  \newcommand{\toc}{}

  }
 \makeatletter\newcommand\mytoc{\@starttoc{toc}}\makeatother 
\long\def\symbolfootnote[#1]#2{\begingroup%
\def\thefootnote{\fnsymbol{footnote}}\footnote[#1]{#2}\endgroup} 

\newcommand{\eqn}[2][]{Equation#1~\ref{eqn:#2}} 
\newcommand{\fig}[2][]{Figure#1~\ref{fig:#2}}

\newcommand{\sect}[2][]{Section#1~\ref{sec:#2}}
\newcommand{\app}[2][]{Appendix#1~\ref{sec:#2}}

\newcommand{\bb}[1]{\ifmmode \mbox{\boldmath $ #1$} \else  \mbox{\boldmath $#1$} \fi}

\newcommand{\mh}{\ensuremath{\textrm{\,--\,}}}    
\newcommand{\ave}[1]{\langle #1 \rangle}          
\newcommand{\U}[1]{\ensuremath{\mathrm{~#1}}}     
\newcommand{\e}[1]{\ensuremath{\times 10^{#1}}}   

\newcommand{\Myr}{\U{Myr}}          \newcommand{\myr}{\Myr}
          
\newcommand{\pc}{\U{pc}}
\newcommand{\kpc}{\U{kpc}}
          
\newcommand{\Msun}{\U{M}_{\odot}}   \newcommand{\msun}{\Msun}

\newcommand{\cc}{\U{cm^{-3}}}

\newcommand{\kms}{\U{km\ s^{-1}}}
\newcommand{\erg}{\U{erg}}

\newcommand{\dex}{\U{dex}}

\newcommand{\mach}{\ensuremath{\mathcal{M}}\xspace}      
\newcommand{\tff}{\ensuremath{t_\mathrm{ff}}\xspace}     
\newcommand{\eff}{\ensuremath{\epsilon_\mathrm{ff}}\xspace}     
\newcommand{\fgas}{\ensuremath{f_{\rm gas}}\xspace}      
\newcommand{\aco}{\ensuremath{\alpha_\mathrm{CO}}\xspace}     
\newcommand{\tdep}{\ensuremath{\tau_\mathrm{dep}}\xspace}        

\newcommand{\ramses}{{\sc Ramses}\xspace}

\newcommand{\magi}{{\sc Magi}\xspace}

\newcommand{\sxb}{Observatoire Astronomique de Strasbourg, Universit\'e de Strasbourg, CNRS UMR 7550, F-67000 Strasbourg, France\label{sxb}}
\newcommand{\usias}{University of Strasbourg Institute for Advanced Study, 5 all\'ee du G\'en\'eral Rouvillois, F-67083 Strasbourg, France\label{usias}}

\newcommand{\lund}{Lund Observatory, Division of Astrophysics, Department of Physics, Lund University, Box 43, SE-221 00 Lund, Sweden\label{lund}}

\newcommand{\chalmers}{Department of Space, Earth and Environment, Chalmers University of Technology, SE-41296 Gothenburg, Sweden\label{chalmers}}

\newcommand{\gs}{\ensuremath{\mathcal{F}10}\xspace}
\newcommand{\gm}{\ensuremath{\mathcal{F}25}\xspace}
\newcommand{\gl}{\ensuremath{\mathcal{F}40}\xspace}

\defcitealias{Renaud2021c}{Paper I}
\defcitealias{vanDonkelaar2022}{Paper II}
\defcitealias{Ejdetjarn2022}{Paper III}

\newcommand{\paperi}{\citetalias{Renaud2021c}\xspace}
\newcommand{\paperii}{\citetalias{vanDonkelaar2022}\xspace}
\newcommand{\paperiii}{\citetalias{Ejdetjarn2022}\xspace}

\begin{document}
\title{From giant clumps to clouds IV: extreme star-forming clumps on top of universal cloud scaling relations in gas-rich galaxies}
\titlerunning{Extreme star-forming clumps}

\author{Florent~Renaud\inst{\ref{sxb},\ref{usias}}\orcidlink{0000-0001-5073-2267} \and 
          Oscar~Agertz\inst{\ref{lund}}\orcidlink{0000-0002-4287-1088} \and
          Alessandro~B.~Romeo\inst{\ref{chalmers}}
          }   
\institute{\sxb\\\email{f.renaud@unistra.fr} \and \usias \and \lund \and \chalmers}

\authorrunning{Renaud et al.}

\date{Received February 24, 2024; accepted March 25, 2024}
\date{}


\abstract{The clumpy nature of gas-rich galaxies at cosmic noon raises the question of universality of the scaling relations and average properties of the star-forming structures. Using controlled simulations of disk galaxies and varying only the gas fraction, we show that the influence of the galactic environments (large-scale turbulence, tides, shear) contributes, together with the different regime of instabilities, to setting a diversity of physical conditions for the formation and evolution of gas clumps from low to high gas fractions. However, the distributions of gas clumps at all gas fractions follow similar scaling relations as Larson's, suggesting the universality of median properties. Yet, we find that the scatter around these relations significantly increases with the gas fraction, allowing for the presence of massive, large, and highly turbulent clouds in gas-rich disks in addition to a more classical population of clouds. Clumps with an excess of mass for their size are slightly denser, more centrally concentrated, and host more abundant and faster star formation. We find that the star formation activity (rate, efficiency, depletion time) correlates much more strongly with the excess of mass than with the mass itself. Our results suggest the existence of universal scaling relations for gas clumps but with redshift-dependent scatters, which calls for deeper and more complete census of the populations of star-forming clumps and young stellar clusters at cosmic noon and beyond.}
\maketitle

\section{Introduction}

Observations of star-forming regions in the Milky Way and nearby disk galaxies have revealed both close-to-universal scaling relations of the cloud properties \citep[e.g.][]{Larson1981, Rosolowsky2003, Rosolowsky2007, Heyer2009, mivilledeschenes2017}, and a clear influence of the galactic dynamics (\citealt{Schinnerer2013, Hughes2013, Sun2020}, also supported theoretically, see e.g. \citealt{Renaud2013b, Meidt2013, Fujimoto2014}), yet with a subtle impact on the star formation efficiency itself \citep{Querejeta2021}. This suggests complex and multiple coupling mechanisms between the scales of the galaxy, the clouds, and star (cluster)-forming knots.

Thanks to pioneer studies with HST \citep[e.g.][]{Elmegreen2007, Guo2012b}, and more recently with ALMA, JWST, and the power of strong gravitational lensing, a wide window is now opened on the gaseous and stellar inner structures of more distant galaxies, in particular at cosmic noon where the cosmic star formation history peaks ($z\approx 1\mh 3$, \citealt{Madau2014}). For instance, with strong lensing, physical resolutions of a few $10 \pc$ can be reached by ALMA at $z=1$ \citep[e.g.][]{Dessauges2023}, and JWST complements such studies by providing integrated properties of the stellar objects \citep[e.g.][]{Claeyssens2023}. Disk galaxies at these redshift host more extreme physical conditions than in the local Universe \citep[e.g.][]{Swinbank2011}, with most of their morphologies being dominated by clumpy structures \citep{Cowie1995, Elmegreen2013b, Guo2015, Shibuya2016, Fisher2017, Sattari2023}, particularly bright in restframe UV \citep{Forster2009, Zanella2015, Cava2018}. In the vast majority of cases and up to very high redshift ($z\approx 7$, \citealt{Treu2023}), such clumpy appearance has an internal (``in situ'') origin \citep[e.g.][]{Forster2009, Tacconi2013, Girard2018}, as opposed to the multi-component structure that galaxy mergers would yield \citep[e.g.][]{Puech2010, Calabro2019}. 

The magnification from lensing, and/or the gain in angular resolution and sensitivity from JWST revealed the hierarchical nature of (at least some of) the young stellar clumps which host sub-structures below the scales of $\sim 1 \kpc$ and $\sim 10^{8\mh9} \msun$ \citep{Mestric2022, Messa2022}, contrarily to previously thought \citep{Elmegreen2005, Forster2011, Guo2012b}. Similar conclusions have also been proposed by adopting observational limitations to detect clumps in simulated galaxies, and finding overestimations by up to a factor of 10 for the clump's stellar masses (\citealt{Huertas2020}, see also \citealt{Faure2021}). These results ask the question of the hierarchical organization of the dense gas and of its star-forming phase in gas-rich clumpy galaxies at $z\approx 1\mh 3$.

Although clumpy galaxies are found in the main sequence of galaxy formation \citep{Schreiber2015, Fisher2019}, the peculiar organization of their interstellar medium (ISM), clustering of star formation, and elevated star formation rate (SFR) influence the distribution and luminosity of the traditional tracers of dense gas and young stars (see e.g. \citealt{Daddi2015} and \citealt{Renaud2019} on the \aco conversion factor in clumpy galaxies), further complicating the interpretation of not-always resolved observational data. Despite the significant recent improvements in precision, completeness, depth, and redshift range, observation programs are still not conclusive on the similarities or differences of star-forming gas structures between nearby galaxies and gas-rich clumpy disks. In particular, it is still not clear whether (and/or at which scale) the structure of the star-forming clouds/clumps and the star formation process are universal across cosmic time, or if new regimes must be invoked in the extreme physical conditions of gas-rich galaxies.

The clumpy nature of gas-rich disks has been well-retrieved in many simulations of galaxies in isolated and cosmological context \citep[among many others, see][]{Noguchi1999, Agertz2009, Dekel2009, Bournaud2010, Ceverino2012, Hopkins2012, Bournaud2014, Tamburello2015, Behrendt2016, Mandelker2017, Fensch2021, Renaud2021c}. This peculiar morphology has been attributed to violent disk instabilities triggered by gas inflows onto an existing disk galaxy \citep{Agertz2009, Dekel2009, Bournaud2010}, or to hierarchical growth by mergers of smaller clumps \citep{Behrendt2016}. In the first paper of this series (\citealt{Renaud2021c}, hereafter \paperi), we show that it is the gas-rich nature of these disks which allows for the development of new regime of instability, replacing the classical Toomre's found at low gas fractions (below 20\% of the baryonic mass of the galaxy). Such instabilities are still driven by the stellar component, but with a significant contribution of the molecular phase in the inter-clump regions. This explains the formation of local clumpy structures, instead of the galactic-wide spiral pattern at lower gas fractions. The highly turbulent nature of these clumps imprints a large velocity dispersion to the young stars, to levels potentially sufficient to explain the kinematics of the thick Galactic disk (\citealt{vanDonkelaar2022}, hereafter \paperii).

While qualitatively retrieved in many simulations of gas-rich disks, the quantitative results on the clump properties vary a lot across the literature. Different numerical techniques, in particular for stellar feedback, lead to different conclusions on fundamental properties of the clumps like their mass \citep{Mandelker2016, Mayer2016} and lifetimes \citep{Genel2012, Bournaud2014, Oklopcic2017, Fensch2021}. Conversely, \citet[hereafter \paperiii]{Ejdetjarn2022} showed that the same level of turbulence could be reached in disk galaxies with and without stellar feedback, thus favoring gravitational instabilities as the main driver of turbulence at galactic scale.

It remains to determine whether the star-forming clumps in gas-rich galaxies are scaled-up versions of the giant molecular clouds in nearby galaxies, requiring a mere extrapolation of the scaling relations established in the local Universe, or if they fundamentally differ in nature. This question of paramount importance is the key to either allow for a transposition to gas-rich disks of our detailed understanding of the physics of the ISM and star formation obtained in multi-wavelength, deep, and highly resolved studies of local galaxies, or on the contrary, to call for new physical regimes specific to the conditions found at high redshift.

In this fourth opus of the \emph{From giant clumps to clouds} series, we use numerical models of isolated disk galaxies, varying only the gas fraction. These simulations lead to spiral galaxies and to clumpy morphologies at low and high gas fractions respectively. The subgrid models adopted here have been used in previous works focusing on the diversity of clouds and the role of feedback in the organization of the ISM, which have concluded to remarkable matches with observations of the ISM of Milky Way-like galaxies \citep{Grisdale2017, Grisdale2018, Grisdale2019}. Here, we examine the properties of gas clumps at various gas fractions, with the aim of testing the hypothesis of universal scaling relations across redshift. 

\section{Method}

\subsection{Simulations}

The simulations and numerical method are introduced in \paperi, and briefly summarized here. We use the \ramses code \citep{Teyssier2002} to model a $7\e{11} \Msun$ galaxy comprising a dark matter, a central bulge, a gaseous and a stellar disk initially setup with the \magi code \citep{Miki2018}. For the sake of control on the parameters, the cosmological context is ignored. Using these initial conditions, we vary the relative mass of the stellar and gaseous components to generate three models, every other parameter being left unchanged. After running the simulations for an initial relaxation phase of $\sim 100\mh 150 \Myr$, the gas fraction $\fgas = M_{\rm gas} / (M_\star + M_{\rm gas})$ in the galaxies is 10\%, 25\% and 40\%. These models are then called \gs, \gm, and \gl respectively.

The simulations reach a resolution of $12 \pc$ and include heating from a cosmological background calibrated at $z=0$ \citep{Haardt1996}, atomic and molecular cooling, star formation at an efficiency (SFE) of 10\% per free-fall time in the gas denser than 100 \cc. The stellar feedback model follows that of \citet{Agertz2021}, i.e. winds, radiation pressure, type-II and Ia supernovae (SNe) and the production of oxygen and iron. In post-processing, the fractions of atomic and molecular gas are computed from cooling tables assuming collisional ionization equilibrium in every cell of the simulation, and following \citet{Krumholz2009}.

As detailed in \paperii, the gas fractions and resulting star formation activity in these runs can be related to that of disk galaxies at different redshifts. An approximate correspondence with the SFRs of Milky Way-like galaxies and supposed progenitors observed in \citet{vanDokkum2013} matches our \gs, \gm, and \gl cases with the redshifts $\approx 0$, $\approx 1.5\mh 2$, and $\approx 2 \mh 2.5$ respectively. We stress however that our controlled experiments are designed to keep the same masses and sizes for all galaxies as we only vary the gas fraction, which implies that our three cases are not representative of different stages along the evolution of the same galaxy.

\subsection{Cloud/clump detection and statistics}

\begin{figure*}
\centering
\includegraphics{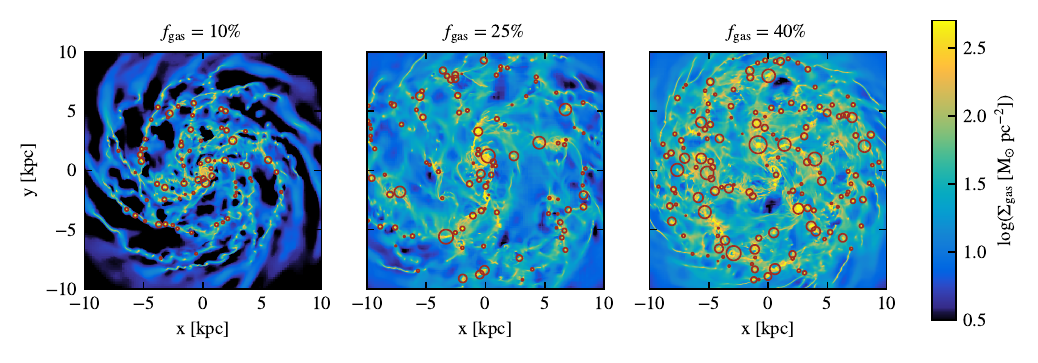}
\caption{Surface density of gas in the three simulations, at the instant considered in the entire analysis. Red circles indicate the position and size of the gas clumps detected (see text).}
\label{fig:maps}
\end{figure*}

Our working resolution of $12 \pc$ is sufficient to detect gas overdensities (clouds and clumps), but not to resolve their inner morphologies and sub-structures. While such fine details are irrelevant for the present paper in the case of giant molecular clouds found at low gas fractions, the situation could be different in much larger and more massive clumps in gas-rich disks, which could host distinct star-forming sub-structures \citep[e.g.][]{Behrendt2016}. By keeping this limitation in mind, we focus on a specific level of the hierarchical ISM, that is the cloud-like gas structures with a high density contrast from their surroundings. Even if our simulations do resolve some of them, sub-clumps are not examined in this work.

We identify gas clumps as over-densities in face-on surface density maps ($> 300 \Msun\, \pc^{-2}$). Their size is then determined by iteratively adding concentric shells centered on the center of mass of the over-density, until the density of gas of a shell is less than 10\% of that of the peak. By doing so, we focus our analysis on the densest structures, and avoid spiral-like elongated shapes. Once this identification is done, the clump is cataloged as a sphere centered on the maximum density.\footnote{We have also tested an option where we stop adding concentric shells when a certain density level, independent of the peak density of the clump, is reached. However, this does not provide satisfactory results: mild over-densities of elongated structures (e.g. spiral arms) would lead to unrealistically extended clumps.} In the case where several density peaks are not separated by a diffuse enough medium and thus overlap, the associated clumps are considered as sub-structures of a single object. As mentioned above, our working resolution is not sufficient to systematically properly capture the formation and evolution of sub-structures. Therefore, the overlapping clumps are merged such that only the largest (containing) one is retained. As a consequence, the center of mass of a clump can be offset from its maximum density. We come back to this point when examining the density profiles of clumps in \sect{profile}.

\begin{figure}
\centering
\includegraphics{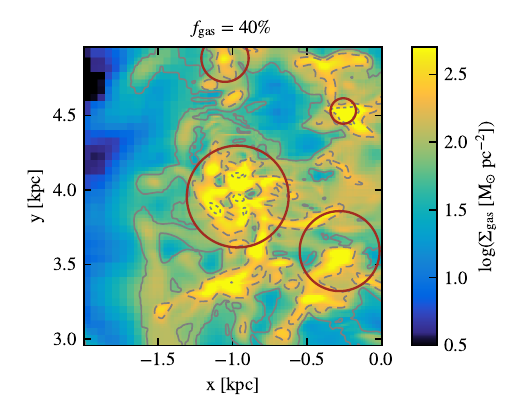}
\caption{Example of a massive clump hosting sub-structures, as a zoom-in on \fig{maps} (right panel, \fgas = 40\%). The contours indicate the surface densities of molecular gas at 10, 100 and $1000 \msun\,\pc^{-2}$. The red circles shows our clump identification. As visible in this example, some of the largest clumps host clear sub-structures, which are inexistent in smaller clouds (either physically or because of limited numerical resolution).}
\label{fig:maps_zoom}
\end{figure}

We have checked that this method provides an identification similar to what a visual examination would lead to, as illustrated in \fig[s]{maps} and \ref{fig:maps_zoom}. In accordance with the objectives of this work, our identification method relies on a structural argument rather than on the gas phase. The main motivation for this choice is to adopt an identification process which works equally well at all gas fractions, without having to adjust parameters between the runs as it could be required by the different morphologies and properties of the gas over-densities. This implies that clumps in our definition do not necessarily coincide with the molecular phase of clouds, but rather with gas overdensities, including their atomic envelopes. Therefore, our clumps are significantly more massive than GMCs in the Milky Way, even in the low gas fraction case. Including the atomic envelope in our selection also provides consistency with the resolution limitations discussed above: molecular sub-clumps of the same structures are merged when they share a (dense enough) atomic envelope. We note that \citet{Grisdale2018} used the same sub-grid recipes as ours and initial conditions comparable to our \gs case to conduct a direct comparison with observed molecular clouds. The good match between their simulations (using mock CO projected maps) and the observations like the power spectrum of the ISM and the clump mass function validates the numerical we also use. This allows us to focus now on physical mechanisms involved in the clumps formation and evolution, in cases where the match with real objects is more difficult to assess. Therefore, we do not adopt a selection method motivated by observations, but rather by physical considerations.

As the bottom of the potential well, the central region of disk galaxies is the convergence point of many gaseous and stellar structures, such that the properties of these structures rather fall under the influence of complex evolutionary processes like accretion and mergers. This makes the interpretation of the clump properties very involved. For this reason, we ignore the gas clumps in this volume, by taking an exclusion zone within a conservative galactic radius of 2 kpc.

This methodology gives approximately 70, 110 and 150 clumps, with average radii of $\approx$ 120, 160 and 180 \pc, for the 10\%, 25\% and 40\% cases respectively. To improve the statistics, this detection is repeated on 5 consecutive snapshots (spanning about $20 \Myr$) for each simulation.

Throughout the paper, we use robust statistics to limit the weight of outliers in our spare datasets. We use the median instead of the mean, we compute the robust standard deviation as the median absolute deviation (MAD) divided by 0.6745, and we perform median-based fits rather then least-squares fits, as in \citet[see their section 2.3 for more information]{Romeo2023}. Furthermore, to robustly identify the outliers of the relation between the measurements $\{x_1,...,x_N\}$ and $\{y_1,...,y_N\}$, we compute the best-fitting relation $y=f(x)$, and the threshold $T$ defined as:
\begin{equation}
	\label{eqn:thres}
	T = \sqrt{2 \ln(N)}\,\frac{{\rm median}\left(\left\{\left|y_i - f(x_i)\right|\right\}^N_{i=1}\right)}{0.6745}, 
\end{equation}
where $N$ is the number of points, and the fraction is the robust standard deviation (see equations 9, 10, and 12 of \citealt{Romeo2004}, for details). A data point $(x_i,y_i)$ is robustly considered as an outlier if $| y_i - f(x_i) | > T$.

\section{Results}

\subsection{Disruptive mechanisms}
\label{sec:disruption}

Different regimes of instability between the low and high gas fractions in Milky Way-mass disk galaxies have been shown to imprint the morphology of dense gas clumps, with a transition between the two regimes occurring at gas fractions of $20 \%$ (\paperi, see also \paperii on kinematics). We examine here whether these differences are amplified or mitigated by environmental effects within and around the clumps.

\begin{figure}
\centering
\includegraphics{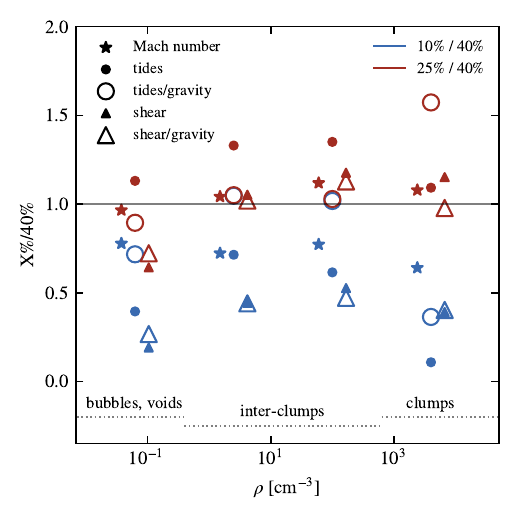}
\caption{Average Mach number (star symbols), and intensity of tidal (circles) and shear (triangles) forces in gas density bins, normalized to that in the $\fgas = 40\%$ case. Empty symbols denote the forces normalized to the local self-gravity. All quantities are measured at a scale of 120 pc. Symbols have been slightly offset horizontally for clarity. The approximate distinctions between three different media are identified at the bottom, for reference. The two most gas-rich cases (red) yield comparable values at all densities for all these quantities, significantly higher than in the low gas fraction galaxy (blue).}
\label{fig:tidesshear}
\end{figure}

\fig{tidesshear} shows, in logarithmic bins of gas density, the intensity of turbulence, tides, and shear, measured at a scale of 120 pc (i.e. over 10 resolution elements, and of the order of the average scale of clumps). The similarities in the gas density probability distribution functions (PDFs, Figure 3 in \paperi) justify using the same density binning for all three gas fractions. The adopted bins approximately encompass the very diffuse ISM found in supernovae bubbles and galactic outskirts ($\sim 10^{-1} \cc$), the inter-arm/inter-clump regions ($\sim 10^1 \cc$), the structures connected to the clumps, like spiral arms ($\sim 10^2 \cc$), and the star-forming clumps themselves ($\gtrsim 10^3 \cc$). The measurements are done on a regular Cartesian grid, without centering on the clumps of any other structure. The tidal force is computed as the maximum eigenvalue of the tidal tensor (i.e. minus the Hessian of the gravitational potential) via first-order finite differences of the gravitational acceleration. Therefore, it is not the azimuthally-averaged tidal strength, but rather the force along the direction where local tides are the strongest (see \citealt{Renaud2011} and \citealt{Renaud2017} for details). The shear force is evaluated as the quadratic sum of the off-diagonal terms of the Jacobian matrix of the velocity field, also computed with first-order finite differences (see \citealt{Renaud2015d}). This quantity can be interpreted as the squared rotational frequency term in the expression of the local centrifugal force. These raw values are rather difficult to interpret and thus we present them in normalized forms, using arbitrarily the case at $\fgas = 40\%$ as reference. 

The turbulence, as traced by the Mach number $\mach$, is very similar between the two gas-rich cases with less than a 10\% difference at any density (as shown by the red star symbols in \fig{tidesshear}). It evolves from the transonic regime ($\mach \approx 0.6$) at the lowest densities, to highly supersonic ($\mach \approx 25$) inside the star-forming clumps. However, the low gas fraction stands apart with a turbulence about 1.5 times weaker in \gs (blue star symbols), with this ratio showing no strong nor clear dependence on the gas density. The fact that gas rich disk galaxies have significantly higher levels of turbulence than those with more modest gas fractions has been well established observationally (see e.g. \citealt{Forster2009, Fisher2017, Girard2019}). Possible explanations invoke the continuous injections of turbulence by internal mechanisms (like feedback from a sustained star formation and kpc-scale disk dynamics, see e.g. \citealt{Hoffmann2012, Agertz2015b, Fisher2019, Brucy2020}), and stirring of the ISM by external factors like gas accretion and more frequent mergers at high redshift \citep{Agertz2021, Renaud2021, Forbes2023}. Using simulations of Milky Way-like progenitor galaxies in isolation, several works reported however that external factors are not necessary to develop and maintain high levels of turbulence in gas-rich disks at redshifts $\approx 1\mh 3$ (\citealt{Clarke2019, Khoperskov2021}, \paperii, \paperiii), and that galaxies evolve intrinsically from strong to moderate levels of turbulence as the gas fraction decreases. Our works show that, all other things being equal, this transition is sharp, with strong turbulence maintained down to gas fractions around $\approx 20\%$, before it drops (see e.g. the Figure 2 of \paperii). Owing that turbulence support is of paramount importance in maintaining a gas disk geometrically thick and dynamically hot (the other main factor being the tidal excitation from satellites and mergers, see \citealt{Renaud2021}), our results imply that, in the absence of mergers, disk galaxies in the mass range we consider undergo a rapid transition from thick to thin disk when their gas fraction reaches $\approx 20\%$. 

Similar trends are also found for the tides and shear, although the differences between the two gas-rich cases are more important than for the Mach number. Yet, they remain within a factor of $\approx 1.3$. Here again, larger differences are found between the \gs and \gl cases: tides are $\approx 1.5 \mh 12$ times weaker and shear is $\approx 2\mh 6$ times weaker at low gas fraction. This further adds to the similarities between the two gas-rich cases, over the entire range of densities in the galaxies, and a sharp transition when moving towards lower gas fractions. This contrasts with the results of \citet{Fensch2021} who reported a shear parameter increasing with decreasing gas fraction. We note however that their parameter does not measure the full shear, but only the differential rotation of the disk via the difference of tangential galactic velocities around clumps. The high velocity dispersions of clumpy disks however implies that the radial and vertical components of the velocity field can also carry non-negligible contributions to the total shear. In addition, the shear parameter of \citet{Fensch2021} does not account for the scale at which it is measured (while ours does, through the transverse velocity gradients). As the characteristic scale of instabilities and structures differs between giant clumps in gas-rich galaxies and spiral arms at low gas fraction (\paperi), these two differences likely explain why our definition of shear leads to opposite conclusions to that of \citet{Fensch2021}. 

Tides and shear result from a complex interplay of phase-space structures, with contributions from both large and small scales, including from the gas clumps themselves. To estimate the effective effects of these forces on the gas structures, we normalize them by the local self-gravity (open symbols in \fig{tidesshear}). This tends to reduce the differences between the galaxies, especially in the inter-clump medium ($\sim 10^1\mh 10^3 \cc$), but the low gas fraction case still stands apart from the other two. The presence of massive clumps is thus an important driver of the phase-space organization in their vicinity (i.e. in the inter-clump medium), via their local gravitational influence (as shown by the tidal force) and other kinematic effects (possibly outflows from feedback interfering with the large-scale disk rotation, as shown by the shear). We come back to this in \sect{profile}.

Overall, \fig{tidesshear} indicates that the main agents capable of disrupting or destroying structures are significantly stronger at high $\fgas$, at all densities. This combines with a higher SFR, and thus a more energetic disruptive effect from stellar feedback. This \emph{could} point toward a shorter lifetime of the clumps at high than low gas fraction. However, the details of the organization of the ISM, the size of the structures and their density profiles, i.e. their fragility to destructive factors, must be accounted for before concluding on the rapid dissolution of clumps. We discuss these aspects in the next sections.

\subsection{Clump mass function}
\label{sec:cmf}

\begin{figure}
\centering
\includegraphics{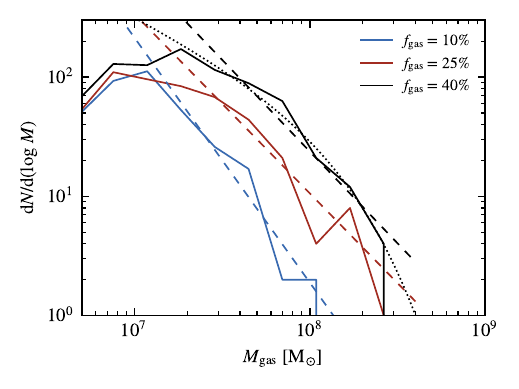}
\caption{Clump mass functions for the three cases (using staking from 5 consecutive snapshots, to improve the statistics). The dashed lines indicate the best-fit of the high-mass end using a power-law. This gives indexes of -2.0, -1.5 and -1.6 for the \gs, \gm, and \gl cases respectively. The dotted curves are fits to the data of \fgas = 40\% using a Schechter function (see text). This functional form does not provide a good match to the other cases.}
\label{fig:cmf}
\end{figure}

The gas density PDFs (Figure 3 of \paperi, reproduced in \fig{pdf_sfefb} below) show important similarities between the three cases, including in the densest phase. In particular, the two gas-rich cases do not host much denser gas than \gs. This naturally implies that the large clumps at high \fgas are not significantly denser, but rather more massive than the clouds at low gas fractions. This is confirmed by the clump mass function shown in \fig{cmf}. Recall that our clump identification method does not probe the low-mass end of this distribution, and we thus focus here on the high-mass end. The maximum clump mass increases with \fgas, leading to a dependence of the high-mass end slope of the mass function with the gas fraction. 

The best-fit power-law relations to the mass functions yields indexes of -2.0, -1.5 and -1.6 for the three cases respectively, all within the range of values observed across galactic environments ($\approx -1.5\mh -2.9$, see e.g. \citealt{Rosolowsky2005}). The mass function of the \gs case is roughly similar to the -2 index well-established in initial star cluster mass functions \citep[e.g.][]{Whitmore1999, Larsen2002, Bik2003, Chandar2014}, and often presented as an outcome of gas fragmentation driven by gravo-turbulence (\citealt{Elmegreen1997}). However, higher gas fractions yield slightly shallower and top-heavy clump mass functions with respect to the low gas fraction, but also in comparison with the observed slopes of -1.7 and -2.0 (depending on the method adopted) in gas-rich disks at $z\approx 1\mh 3.5$ \citep{Dessauges2018}. This difference means that gas-rich disks do not simply host massive clumps in addition to those found at lower \fgas, but that they also count more numerous clumps at intermediate masses. The shapes of the mass functions of the two gas-rich cases are very similar, but they differ by the overall number of clumps.

We note that the mass function from the \gl run shows a remarkable match with a \citet{Schechter1976} function of the form $\propto M_{\rm gas}^{-0.73} {\rm exp}[-M_{\rm gas}/(1.3\e{8} \msun)]$ (dotted line in \fig{cmf}), while this is not the case for the other galaxies. Some studies proposed that the initial mass function of star clusters follows a Schechter form \citep{Larsen2009, Adamo2017}, which could suggest a direct mapping between our clumps and observed clusters. However, the presence of an exponential cutoff differencing the Schechter functional form from the classical power-law could have a physical or a statistical origin. As discussed in \citet{Mok2020}, small datasets could induce an apparent cutoff in the mass function, simply because of the statistical lack of objects to populate the high-mass end of the distribution. This would then differ from a cutoff imposed by physical mechanisms. Our analysis suffers from the same problem, and we are not able to conclude on the origin of the exponential cutoff. Yet, it is noteworthy that only the galaxy with the highest gas fraction (which is also the one hosting the most clumps and thus which is potentially the least sensitive to a statistical effect on the cutoff) is the only case yielding an agreement with the Schechter functional. If physically-driven, the reason for a Schechter mass function only in the gas-rich case is difficult to identify. It could result from a truncation of the most massive clumps, possibly by stellar feedback and/or environmental effects (\sect{disruption}). \citet{Hopkins2012} proposed an analytical model retrieving the exponential cutoff at the high-mass end of the mass function of GMCs. We note however that this is a direct consequence of the assumptions made on the shape of the power spectrum of the ISM, thus leaving open the question of the fundamental cause(s) of such a cutoff (if physical). From our experiments, it is not clear whether the truncation is intrinsic to massive clumps (regardless of other properties), or rather to the high gas-fraction. To investigate these hypotheses, the next section focuses on the structure of the most massive clumps in our galaxies.

To examine whether our conclusions on the influence of the gas fraction on the clump mass function is robust to a change of the subgrid recipes employed, we show in \app{subgrid} examples of the same analysis conducted with different parameters and ingredients for the star formation and feedback models. We also show the density PDFs, for reference. The conclusion is that, within the relatively wide range of parameters explored, changing the subgrid models does not alter the clump mass function as strongly as the gas fraction does.

\subsection{Clump properties}

\subsubsection{Clump density profile}
\label{sec:profile}

\begin{figure}
\centering
\includegraphics{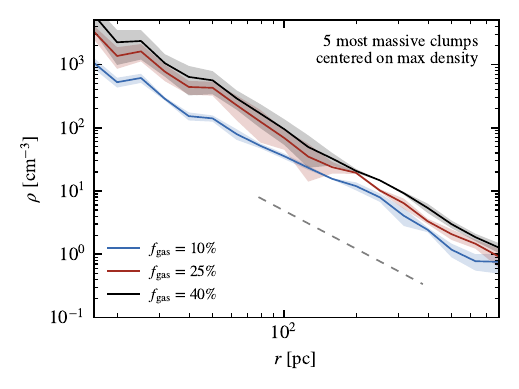}
\includegraphics{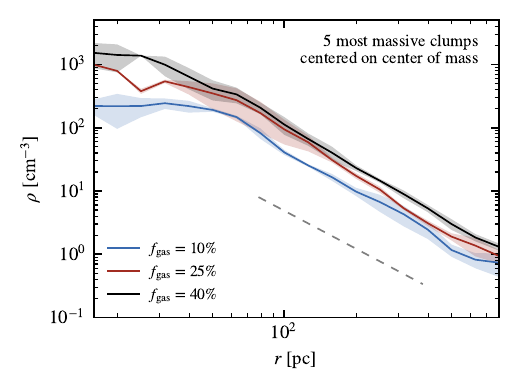}
\caption{Median density profile of the five most massive gas clumps in each simulation (excluding the central 2 kpc of the galaxies, where radial inflows at galactic scale play a non-negligible role in the accumulation of gas). The top panel shows the radial profiles with the maximum density as the center, and the bottom panel those centered on the center of mass. The shaded areas show the robust standard deviation over these 5 clumps. The dashed line represents a power-law of index -2, for reference.}
\label{fig:clumpprofile}
\end{figure}

\fig{clumpprofile} shows the gas density radial profiles of the five most massive clumps in our three galaxies (excluding the central-most, bulge-like, overdensities). Two sets of profiles are computed, using the maximum density and the center of mass as centers. The difference originates from the finite resolution of the simulations\footnote{The maximum density necessarily corresponds to the center of a simulation cell, contrary to the center of mass which is an average position. This difference influences the innermost parts of the density profile, close to the resolution limit ($12 \pc$).}, and the hierarchical structure of clumps, as discussed below (also recall \fig{maps_zoom}). In the following, we ignore the central parts of the profiles (up to $\approx 30\mh 40 \pc$), affected by the resolution limits. Potential sub-structures in these clumps do not necessarily exist at the same distance from the center of each clump, and thus their contributions are rapidly averaged out in the median radial profiles shown here. In this section, we ignore the size of the clumps and show the radial profiles from the clump centers up to large distances.

Once again, the cases at $\fgas = 25\%$ and $40\%$ are remarkably similar, while the \gs case stands apart. Clumps in the two gas-rich cases are slightly but significantly denser than those in the \gs galaxy by factors ranging from 1.5 to 3.1, up to several $100 \pc$, i.e. where reaching the inter-clump medium. However, the density of clumps from the two gas-rich disks are remarkably similar at all radii. Therefore, it is the organization of the ISM, initiated by the instabilities discussed in \paperi\footnote{Because of the non-linearity of the growth of structures, it is not possible to directly connect the sizes of clumps to the characteristic scales of \emph{any} stability analysis (like Jeans' or Toomre's).}, and not (directly) the increase of the gas mass of the disk which is responsible for the increased density of the most massive clumps in gas-rich galaxies.

When using the maximum density as the center of the clumps, the clumps profiles exhibit power-laws slightly steeper in the gas-rich disks ($\approx -2.7 \mh -2.2$, depending on radius) than at low gas fraction ($\approx -1.8 \mh -1.5$), but always close to the -2 index found for an isothermal sphere.

When centering on the center of mass, the presence of sub-structures near but not exactly at the center creates density peaks at various radii and azimuths, which average out as a flattening of the density profiles as illustrated in \app{exprofiles}. The central cores in \fig{clumpprofile} are thus a statistical effect, and are generically not representative of the actual profile of a given clump taken alone.

\begin{figure}
\centering
\includegraphics{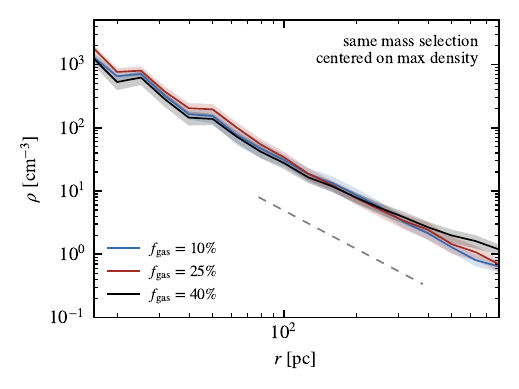}
\includegraphics{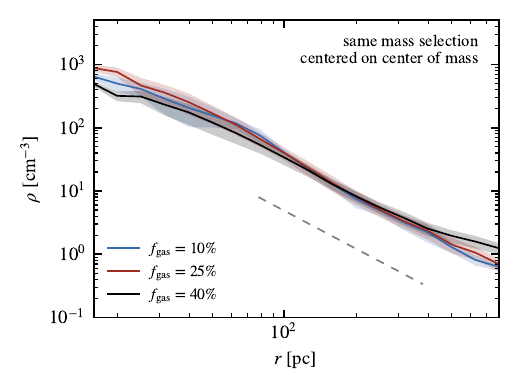}
\caption{Median density profile of gas clumps in each simulation (as in \fig{clumpprofile}), but only considering clumps in the same mass interval. Clumps of comparable masses have similar density profiles regardless of the gas-fraction of their host galaxy.}
\label{fig:clumpprofilemass}
\end{figure}

By selecting the most massive clumps of each simulation, we introduce a potential bias because the high-mass end of the clumps mass functions of the three simulations do not overlap (recall \fig{cmf}). Yet, it is not obvious whether this difference in the clump mass is the sole explanation for the differences in the density profiles. \fig{clumpprofilemass} confirms this hypothesis by showing that the density profiles of clumps selected in a common mass range populated in all three galaxies ($2\e{7} \mh 1\e{8} \msun$) are remarkably similar. In fact, the differences between the three cases are even smaller than the clump-to-clump scatter within a given simulation (shown as shaded areas). Therefore, \emph{at a given mass}, a gas clump has a density profile independent of the gas fraction of its host disk. The different profiles only arise because of the presence of more massive clumps in gas-rich galaxies. As explained above, these more massive cases are denser, and possibly yield a central core. (This conclusion could be affected by our resolution limitations.) Therefore, such clumps ought to be more resistant to external disruptions (\sect{disruption}), and internal effects like feedback. This suggests that massive clumps in gas-rich galaxies can be longer-lived than local GMCs, which supports, at least qualitatively, the conclusions of previous works on clump survival over several orbital times ($\gtrsim 100 \Myr$, see e.g. \citealt{Bournaud2015} and \citealt{Dekel2023}, but see \citealt{Genel2012} and \citealt{Oklopcic2017} for opposite findings).

Our results illustrate the need for capturing the inner structure of the clumps and the spatial (and time) scatter of sources of feedback simultaneously with galactic dynamics to definitely conclude this long-lasting debate.

\subsubsection{Larson scaling relations}
\label{sec:larson}

\begin{figure*}
\centering
\includegraphics{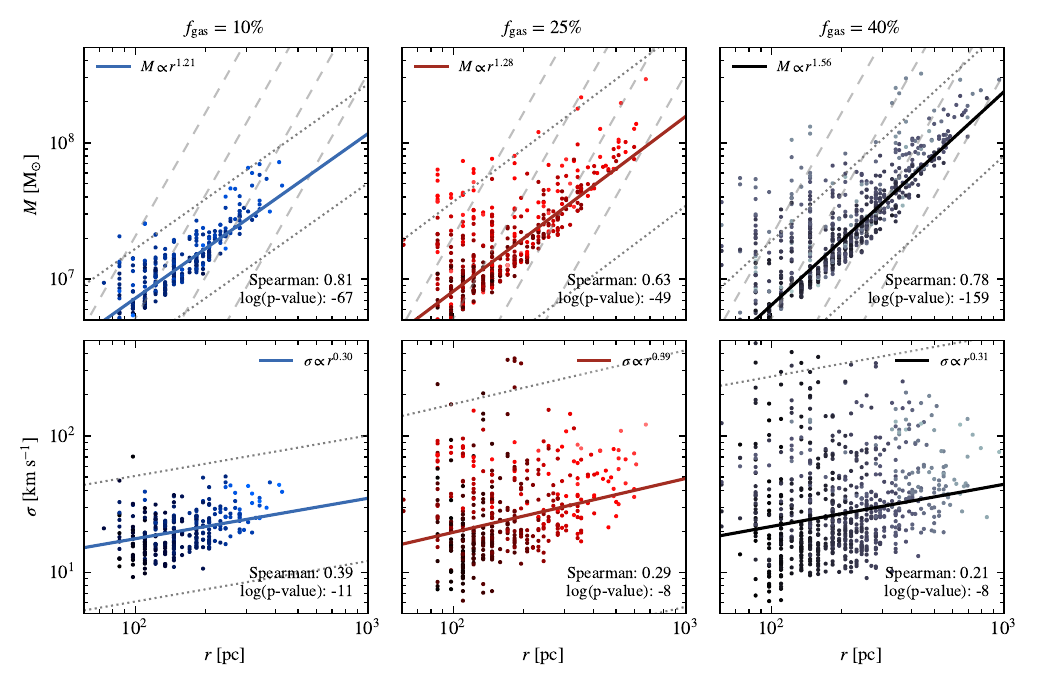}
\caption{Mass-size (top) and velocity dispersion-size (bottom) relations for the clumps in our simulations. The color of the points encodes the logarithm of the third quantity (velocity dispersion on the top panels, and mass in the bottom row), increasing from dark to light. The solid lines are the least absolute deviation fittings of the linear relations in log-log space. They are translated into a power-law in linear-linear space, with the indexes indicated at the top of each panel. The dashed lines show constant densities, for reference, and the dotted lines mark the threshold for the robust identification of outliers (see text, \eqn{thres}).}
\label{fig:larson}
\end{figure*}

In the previous section, we found that some properties of gas clumps like their density profiles are independent of the gas fraction of their host galaxy, at a given clump mass, but that a different regime of instability in gas-rich disks allow for the formation of more massive and denser clumps (\paperi). This duality questions the universality of scaling relations for gas clouds, which we examine here.

\fig{larson} shows the distributions of all the clumps detected in the simulation, over five consecutive snapshots to improve the statistics, in mass $M$, radius $r$, and velocity dispersion $\sigma$. Observational census of molecular clouds in the Milky Way \citep[e.g.][]{Falgarone2009, Heyer2009, mivilledeschenes2017}, nearby galaxies \citep[e.g.][]{Hughes2013}, and some examples are high redshift \citep{Dessauges2023} report close to universal relations between these quantities, as first noticed by \citet{Larson1981}. Recent estimates of these so-called Larson relations in the Milky Way are $M\propto r^{2.2 \pm 0.2}$ and $\sigma \propto r^{0.63\pm 0.30}$ \citep{mivilledeschenes2017}. As noted above, our method does not only detect the molecular contents of clumps, but also their atomic envelopes around the density peaks. Indeed, the diversity of clump sizes and morphologies between low and high gas fractions, and the presence of sub-structures in some of the most massive clumps (\fig{maps_zoom}) imply that the molecular phase alone does not necessarily cover the entirety of a clump structure, specially in gas-rich disks. The consequence of this selection criteria is that, compared to molecular clouds, our clumps encompass a larger amount of gas at intermediate densities. Therefore, the mass-size relation of our clumps is necessarily shallower than that of molecular clouds, as confirmed by the fits in \fig{larson}. These fits are computed in log-log space as least absolute deviation fittings, i.e. linear regressions similar to the least squares technique, but using absolute values instead of square values to reduce the influence of outliers. We perform this computation following the {\tt medfit} algorithm of \citet{Press1992}. 

We indicate in \fig{larson} the Spearman coefficient and its p-value for each relation.\footnote{The Spearman coefficient is computed as the covariance of the ranks of the quantities, divided by the product of their standard deviations. It indicates correlations (resp. anti-correlations) when close to unity (resp. -1). By using rank instead of raw variables, it differs from the more commonly used Pearson coefficient by allowing for the detection of monotonous, and not only linear, relations. Small p-values indicate significant correlations or anti-correlations.} For the three gas fractions, the mass and size of clumps yield relatively strong correlations of high significance. However, the correlations are much weaker for the velocity dispersion-size distributions, due to an important scatter which increases with the gas fraction. This scatter mainly originates from clumps having high velocity dispersions, interestingly at all masses and sizes. When reported to the escape velocity (estimated roughly by using the total mass and radius as $\sqrt{2GM/r}$), the high $\sigma$'s indicate that these clumps are the least bound, and on the verge of dissolving. We note that the outliers in the $r$-$\sigma$ relation are not necessarily the same as those in the $r$-$M$ relation. Therefore, the elevated velocity dispersion is not, or at least not always, a consequence of a mass higher than average.

A possible reason for an elevated velocity dispersion in a clump is the injection of energy and momentum by stellar feedback. When examining the indicators of recent star formation activities in these clumps (star formation rate, efficiency, and depletion time, see \sect{sf}), we find that they cover a large range of values, not exclusive to the high-$\sigma$ clumps. Thus, stellar feedback cannot be the sole factor of the high levels of velocity dispersion for all these clumps. As already noted in \sect{disruption}, (hydro)dynamical disruptive effects from the galactic environments are stronger at high gas fractions, which probably plays an important role, with feedback, in creating the scatter in the distributions of velocity dispersion. Telling apart the external (tides, shear, large-scale turbulence) and internal (feedback) effects on a clump-by-clump basis requires dedicated experiments, out of the scope of this paper. Nevertheless, we examine the star formation activity in the extreme clumps in the next section.

As hinted in \sect{profile}, clumps in the gas-rich disks are not necessarily denser than their counterparts of the same mass at low gas fraction. The similarity of the best fit for the Larson's relations between the three cases indicates in fact that the denser clumps, outliers of the mass-size relation which are more numerous at high gas fraction, are not necessarily the most extended nor the most massive.

The PHANGS-ALMA program performed a census of molecular clouds in nearby disk galaxies and examined their mass-size and velocity dispersion-size relations across a diversity of galactic environments \citep{Rosolowsky2021}. They reported similar average relations as for the Milky Way clouds, but with significantly elevated scatter, which they attribute
to both technical limitations (e.g. resolution bias and source blending) and a diversity of physical conditions (e.g. external forces).

At higher redshift, \citet{Dessauges2019} and \citet{Dessauges2023} combined the high angular resolution of ALMA with strong lensing to detect massive clumps in gas-rich galaxies at redshift $\approx 1$. When examining the properties of the massive molecular clumps, they found an offset with respect to the Larson's relations from local GMCs, in a similar fashion as clouds in local starbursting galaxies: these clumps tend to be more massive, denser, and with a higher velocity dispersion than local GMCs. These results also show an increased scatter in the two scaling relations. This led the authors to question the universality of the gas clumps scaling relations, and to confirmed the prime importance of the galactic environment in setting up the properties of clouds, which had been previously noted observationally \citep[e.g.][]{Swinbank2011, Hughes2013, Sun2020} and theoretically \citep[e.g.][]{Renaud2013b, Grisdale2017}.

Our results are compatible with the findings of \citet{Dessauges2019} and \citet{Dessauges2023}, but we nuance their interpretation: the census of gas structures in distant galaxies is biased toward dense and large objects which overcome the resolution and signal-to-noise limits imposed by the instrument, the gravitational lensing, and the methodology adopted. Therefore, it is likely that the smallest and least dense clumps are not detected, or do not satisfy the selection criteria. In our analysis, the densest and, to a lower extent, the largest clumps are precisely the ones the furthest from the median mass-size relation (top row of \fig{larson}), and hence the main contributors to the increasing scatter of this relation with the gas fraction. Considering these extreme clumps \emph{only} would then, indeed, lead to the conclusion that local scaling relations break in gas-rich galaxies.

From our results, the emerging picture is instead that of a rather universality in the \emph{average} properties and scaling relations of gas clumps (mass, size, velocity dispersion, density profile), together with a scatter in these properties which significantly increases with the gas fraction of the host galaxy, and allows for more frequent extreme outliers. Important information on the variations of the star-forming regions with the galactic environment and also their evolution with cosmic time could then be revealed from the study of these outliers, as we show next.

\subsubsection{Extreme clumps and star formation activity}
\label{sec:sf}

\fig{larson} shows that the three galaxies yield comparable scaling relations, including in terms of intercept or normalization. The main difference resides in the scatter which considerably increases with the gas fraction. Gas-rich disks thus provide the physical conditions allowing some of their gas clumps to depart from the median scaling relations and to reach extreme properties, at formation and/or after some dynamical evolution. To investigate the underlying reasons behind this scatter, we quantify the deviation of a clump from the best-fit relations plotted in \fig{larson}, as the excess (or deficit) each clump has with respect to the relation. For instance, we compute the excess mass $\Delta_M$ of a clump of mass $M$ and radius $r$ as
\begin{equation}
\Delta_M = \log\left(\frac{M}{b\, r^a}\right),
\end{equation}
where $a$ and $b$ are the best-fit parameters of the mass-size relation $M = b\, r^a$. In other words, $\Delta_M$ is the vertical distance between a point and the line in the top panels of \fig{larson}. It is counted negatively for a deficit of mass with respect to the best-fit relation. We compute the excess mass $\Delta_M$ and excess size $\Delta_r$ using the mass-size relations, and the excess velocity dispersion $\Delta_\sigma$ using the velocity dispersion-size relations, but studying $\Delta_M$ provides the strongest conclusions.

\begin{figure*}
\centering
\includegraphics{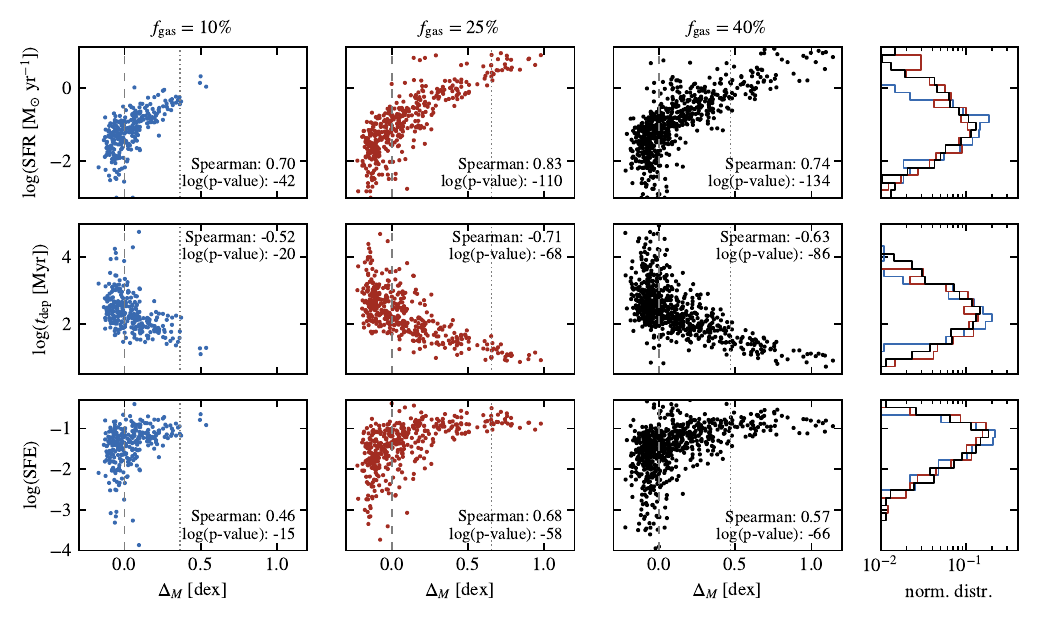}
\caption{Relation between the star formation rate (top), depletion time (middle), and star formation efficiency (bottom), and the excess of mass of clumps with respect to the best-fit mass-size relations of \fig{larson}. The vertical dashed lines mark $\Delta_M = 0$, i.e. a clump sitting on the best fit relation shown in \fig{larson}. The vertical dotted lines show the thresholds for the robust definition of outliers (see text). Normalized marginal distributions are shown on the right column.}
\label{fig:excessmass}
\end{figure*}

\fig{excessmass} shows the excess mass $\Delta_M$ with respect to the star formation rate (SFR), the depletion time (\tdep), and the star formation efficiency per free-fall time (SFE, $\eff$) for each clump. To measure these quantities, we consider the timescale for detection of star clusters via free-free emission $t_y$ chosen equal to $4 \Myr$, and the mass $M_{\rm \star, y}$ of stars younger than $t_y$ in the clump (see \citealt{Grisdale2019} for details). This gives:
\begin{equation}
	{\rm SFR} = \frac{M_{\star, y}}{t_y}, \qquad 	\tdep = M \frac{t_y}{M_{\star, y}}, \qquad \eff = \frac{\tff}{t_y}\frac{M_{\star,y}}{M_{\star,y} + M},
\end{equation}
using the average free-fall time of the clump: $t_{\rm ff} = \sqrt{4\pi^2 r^3 / (32 G M)}$. 

For reference, we indicate in \fig[s]{larson} and \ref{fig:excessmass} the thresholds defining outliers of the relations (recall \eqn{thres}). For instance, a clump with excess mass above $T$ is statistically robustly identified as an outlier of the mass-size distribution. This leads to 1.7\%, 4.8\%, and 8.2\% of the clumps to be classified as outliers in mass, for the \gs, \gm, and \gl cases respectively, all having a positive excess of mass ($\Delta_M >0$).

The normalized distributions of the SFR (top-right panel in \fig{excessmass}) reveal, once again, the similarities between the two gas-rich cases which share comparable distributions and ranges of values. At low gas fraction however, only the low-SFR end of the distribution matches that of the other galaxies. No clump in \gs reaches SFRs as high as the extreme cases in the \gm and \gl cases. These high values are found in the clumps with the highest excess of mass, which are more numerous and further from average in the gas-rich disks. An analogy with the main sequence of star formation at galactic scale would suggest that the highest SFRs are found in the most massive clumps \citep[e.g.][]{Tacconi2018}. We show in \app{masssfr} that this is not the case: a much stronger correlation is found between the SFR and the excess of mass than with the clump mass itself. At first order, this can be interpreted as a result of elevated density at high $\Delta_M$, but the complex structure of clumps and regulation effects from feedback likely complicate this picture. We thus present further analysis, using other indicators of the star formation activity.

\begin{figure*}
\centering
\includegraphics{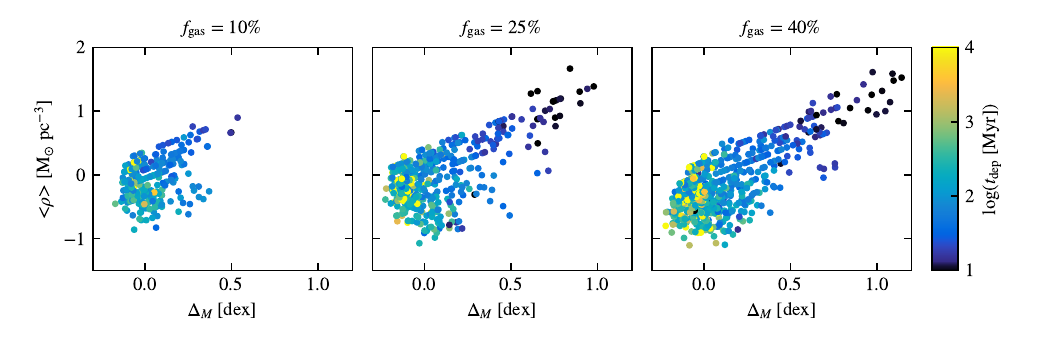}
\caption{Distributions of average density (total mass divided by total volume) and excess mass of the clumps, with color indicating the depletion time. Correlations between these three quantities are only found for the extreme clumps at high $\Delta_M$.}
\label{fig:excessdensity}
\end{figure*}

The middle row of \fig{excessmass} shows that all the clumps at high $\Delta_M$ yield short \tdep, i.e. fast star formation, regardless of the galactic gas fraction. However, the small scatter of the mass-size relation at low gas fraction implies that only a few such clumps exist in the \gs case. \fig{excessdensity} complements this analysis by showing that the most extreme clumps ($\Delta_M \gtrsim 0.8$ and $\log{(\tdep / 1 \Myr)} \lesssim 1.2$) are also those with the highest average density $\ave{\rho} = M / (4 \pi r^3 / 3)$. However, clumps with moderate excess mass ($\Delta_M \approx 0.3\mh 0.8 \dex$) still have short depletion times (a few $10 \myr$), but yield a wide range of densities ($\ave{\rho} \sim 0.1 \mh 10 \msun\, \pc^{-3}$). 

\begin{figure*}
\centering
\includegraphics{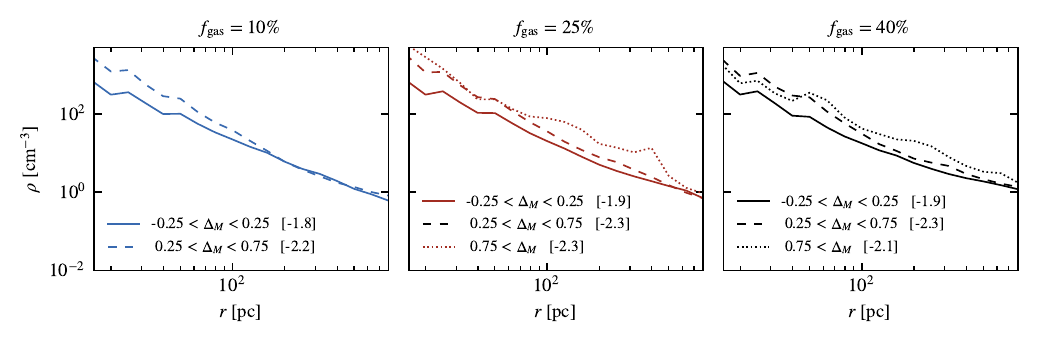}
\caption{Median density radial profiles of clumps of radius between $2\e{7} \msun$ and $3\e{7} \msun$, grouped in three bins of excess mass $\Delta_M$. The numbers in square brackets indicate the slope of each curve measured in log-log space in the inner $100 \pc$. (The conclusions do not change when considering larger radial ranges.) Higher excess masses correspond to more centrally concentrated clumps (steeper profiles).}
\label{fig:profilesclumps}
\end{figure*}

Examining the density radial profiles of individual clumps suggests that the contrast between the density in the center and the outskirts increases with $\Delta_M$. Doing this analysis systematically is difficult because of the dual effect of $\Delta_M$ and $M$ itself on the density profile. To illustrate this, we select clumps of a fixed mass, and study the influence of their excess mass on their density profiles. We choose to select all the clumps in the mass interval $2\e{7}\Msun < M < 3\e{7} \msun$. This interval is large enough to contain several clumps in all the three simulations and thus limit the statistical noise, and narrow enough to restrict the analysis to a precise mass. We then group clumps in three bins of $\Delta_M$: close to average ($-0.25 < \Delta_M < 0.25$), intermediate ($0.25 < \Delta_M < 0.75$), and extreme ($0.75 < \Delta_M$), and compute the median of the density profiles in each bin. The results are displayed in \fig{profilesclumps}. Note that clumps are less numerous in bins of high $\Delta_M$, and thus the profiles are more subject to statistical fluctuations. Nevertheless, clumps with an excess mass ($\Delta_M > 0.25$) yield steeper density profiles than those closer to the scaling relations ($\Delta_M \approx 0$), meaning that such extreme clumps tend to be more centrally concentrated, rather than only simply denser on average (as the sole information of their excess mass for their respective size could indicate). The high peak density in extreme clumps can be connected to their short depletion times. In fact, previous works have shown that reduced depletion times results from an excess of dense gas at galactic \citep{renaud2014b, segovia2022} and cloud scales \citep{renaud2019b}, triggered by tidal and turbulent compression in galaxy mergers. For the present study of disks in isolation, we find that elevated $\Delta_M$'s correspond to steep density profiles, and short depletion times, in line with the conclusions in galaxy mergers. The source of the compression of clumps in our isolated galaxies is not clear, but is likely an outcome of the instability regime highlighted in \paperi. The non-linearity of the growth of these instabilities, and their collapse leading to star formation make it difficult to push this investigation further without dedicated and controlled experiments at high resolution.

The distribution of depletion times (right column of \fig{excessmass}) is slightly narrower at low gas fraction, due to the lack of clumps with extreme mass excesses (as discussed above), but also the rarity of average clumps with very long depletion times. The two cases at high gas fraction yield very similar distributions of \tdep.

Finally, the bottom row of \fig{excessmass} shows the distribution of SFEs. Clumps with an excess of mass yield only high SFEs, while those closer to average display a large range of efficiencies, over several orders of magnitude. Contrary to the SFR and the depletion time, the distributions of SFEs are very similar for the three gas fractions. These results support the conclusions of \citet{Grisdale2019} of a wide diversity of efficiencies in clouds, with no clear correlation with the cloud properties. As found in \citet{Grisdale2019}, the ad hoc value of the SFE per free-fall time implemented in the subgrid recipe for star formation (10\% here) is not representative of the SFE at cloud-scale. Part of this discrepancy originates from the range of densities (and thus local free-fall times) found within a given clump. This range tends to be limited in the densest cases, which could explain the peak of the SFE distribution around 10\% in \fig{excessdensity}, especially at high $\Delta_M$. However, other factors must be taken into account to explain the scatter in SFE, like the fact that clumps are found (and observed) at various stages of their star formation histories, and thus with evolving stellar-to-gas mass ratios. In addition, \citet{Grisdale2019} reported that the stellar feedback and galactic dynamics make the evolution of the SFE of clumps non-monotonic but rather complex, further increasing the diversity across a galactic disk. Some of the clumps in our gas rich disks undergo erosion and mergers with their neighbors, usually in a relatively dense background environment, which makes their tracking in time very difficult. We have not been able to systematically follow the evolution of our clumps, and thus leave these questions for future works.

For all three gas fractions, the excess velocity dispersion does not show strong nor significant correlations with the star formation indicators (SFR, \tdep, SFE), as illustrated in \fig{excessvelocity}. Such a relation would be expected if stellar feedback would be the main source of turbulence within the clumps. However, galactic dynamics play an important role in the setting the turbulence cascade and thus the velocity dispersion in clumps \citep[e.g.][]{Hoffmann2012, Renaud2013b,Ginzburg2022}, which blurs the effects of feedback. Indeed, \fig{larson} already hinted that the excess mass does not correlate with the excess velocity dispersion (the darkness of the points does not correspond to their vertical distance to the solid line). This suggests that different factors set the mass and the star formation activity on the one hand, and the size and large-scale velocity dispersion of the clumps on the other. 


\section{Conclusion}

Using controlled simulations of isolated disk galaxies, we have studied how the gas fraction influences the properties of the star-forming gas clumps, and whether universal scaling relations could be identified. Our main conclusions are as follows.
\begin{itemize}
\item Destructive factors like turbulence, tides, and shear are significantly stronger in gas-rich galaxies than at low gas fractions, at all densities of the ISM. The sharp transition found in the instability regime (\paperi) and in the kinematics (\paperii) at $\fgas \approx 20\%$ thus also occurs in the strength of the destructive effects.
\item The mass function of gas clumps is slightly shallower at high gas fraction, due to the presence of more massive clumps in gas-rich galaxies. The presence of a Schechter-like exponential cutoff is possible at high \fgas, although subject to statistical effects.
\item The radial density profile of a gas clump of a given mass is independent of the gas fraction of its parent galaxy. However, gas-rich disks ($\fgas > 20\%$) host more massive clumps, which are more centrally concentrated (steeper profiles).
\item The scaling relations between the mass, size and velocity dispersion of clumps yield the same slope and intercept at all gas fractions, in a similar way as Larson relations. However, the scatter of clumps around the median relations significantly increases with the gas fraction. In other words, \emph{on average} and at a given mass, clumps have essentially the same properties regardless of the gas fraction, but gas-rich galaxies allow for the formation of extreme clumps not found at low gas fractions, which causes the increased scatter of the relations. Outliers are more massive, more turbulent and tend to be denser than average, which could bias the observational census of clump populations.
\item The excess of mass of a clump above the median scaling relation correlates with a high SFR and a short \tdep, i.e. abundant and fast star formation. The most extreme cases are found in the most centrally concentrated clumps, i.e. with steep density profiles. Such extreme star formation activity correlates much more strongly with the excess of mass for a given size, than with the mass itself.
\item Clumps close to the scaling relation yield a diversity of SFEs, with no clear link with their properties. However, only high SFEs are found in outliers to the scaling relations, with large mass excesses.
\item The variations of the gas fractions have a stronger impact on the properties of clumps and of the ISM than changing the sub-grid recipes for star formation and feedback.
\end{itemize}

For virtually all the quantities and processes studied in this series of papers, we found remarkable similarities between all galaxies with gas fractions $> 20\%$. This suggests an independence of these quantities and processes with redshift for disk galaxies above $z \gtrsim 1$. We suspect however that, if confirmed, this statement might break at higher redshifts ($z \gtrsim 5\mh 6$), when the galaxy is not yet rotation-supported \citep{Segovia2022} and thus when other instabilities are probably active, and when the gas contents of small structures and their star formation activity might be affected by re-ionization.

Our results suggest that the properties of star formation (rate, efficiency, depletion time) span large ranges of values with weak or no correlations with the intrinsic properties of the clumps. However, stronger correlations appear for the clumps more massive than average for their respective size. Such clumps are more frequently found in gas-rich galaxies due to a special instability regime favoring their formation (\paperi). This regime influences the environmental factors acting on the clumps (tides, shear, large-scale turbulence), independently of the clump's intrinsic properties.

The rapid dynamical evolution of clumps which repeatedly merge and split makes it very difficult to track them in a self-consistent manner. This is why we have refrained from tackling the long-lasting question of survivability and lifetime of the massive gas clumps. Nevertheless, we propose here some considerations, to be taken with caution. \citet{Dessauges2023} found separations of a few $100 \pc$ between their detected gas clumps (identified as above a signal-to-noise threshold in CO (4-3) emission) and young star clusters ($10\mh 100 \myr$, with large uncertainties). They then argued that such separations would require an important drift of the stars if they would have formed in the clumps, and that the stellar ages would rather suggest that their natal gaseous nursery would have already been dispersed by feedback (owing a short lifetime for GMCs in nearby galaxies, $\approx 10 \Myr$). Keeping in mind that our definition of clumps largely differs from that of \citet{Dessauges2023}, we note nevertheless that the massive clumps in our gas-rich cases yield molecular sub-clumps separated by a few $100 \pc$ (recall \fig{maps_zoom}, see also \citealt{Faure2021}), and with velocity dispersions\footnote{Note that this velocity dispersion would be significantly smaller if considering the densest star-forming phase of the clumps only, as in \citet{Dessauges2023} who reported estimates of $5\mh 25 \kms$.} up to several $100 \kms$ at these scales (\fig{larson}). Therefore, it is possible that sub-structures within a massive clump could be separated by several $100 \pc$, either intrinsically, and/or after a dynamical evolution of a few Myr. This raises a cautionary note on using physical separations between gas and stellar structures to infer the lifetimes of clumps in gas-rich galaxies.

The denser and possibly cored structure of the extremely massive clumps found in our gas-rich galaxies (\fig{clumpprofile}) make it possible for these clouds to be long-lived. Clumps closer to the average scaling relation, and thus with properties comparable to the clouds at low gas fractions, would be as short-lived as GMCs in the present-day Milky Way, or even shorter-lived due to stronger disruptive mechanisms (\sect{disruption}). Therefore, our conclusions support the idea of two regimes of cloud lifetimes proposed by \citet{Dekel2023}, with the long-lived regime only existing for the most massive clumps in gas-rich galaxies. Reaching more affirmative conclusions on this matter is made difficult by the evolving nature of gaseous structures, an aspect that should be taken into account in future studies.

\section*{Acknowledgements}
We thank the reviewer for comments that helped improve the clarity of the paper. We thank Marc Huertas-Company, Avishai Dekel, and Joel Primack for fruitful discussions. FR acknowledges support provided by the University of Strasbourg Institute for Advanced Study (USIAS), within the French national programme Investment for the Future (Excellence Initiative) IdEx-Unistra. OA acknowledges support from the Knut and Alice Wallenberg Foundation, the Swedish Research Council (grant 2019-04659) and the Swedish National Space Agency (SNSA Dnr 2023-00164).

\bibliographystyle{aa}
\bibliography{biblio}

\appendix
\section{Robustness to changes of subgrid models}
\label{sec:subgrid}

\subsection{Method}

To test the influence of the subgrid models used for star formation and feedback on our conclusions, we restarted the simulations presented in the main text while varying one parameter or ingredient at a time:
\begin{itemize}
\item high SFE: the SFE used in the star formation model is multiplied by 3 (from 10\% to 30\% per free fall time),
\item low SFE: the SFE used in the star formation model is divided by 10 (from 10\% to 1\% per free fall time),
\item energetic SNe: the energy of individual supernova is multiplied by 10 (from $10^{51} \erg$ to $10^{52} \erg$),
\item no pre-SN: the stellar feedback only comprises the type-II and type-Ia SNe (winds and radiation pressure are turned off),
\end{itemize}
which leads to 3 gas fractions $\times$ 5 modifications = 15 runs in total. 

To make the comparisons as direct as possible with the fiducial runs, these additional runs are started from an evolved snapshot of the fiducial set, and restarted with the new parameters for 30 \Myr. This duration ensures that the modifications effectively impact the stellar feedback budget, at least at cloud scale. Capturing a full baryonic cycle would require waiting longer for the re-cooling, re-condensation, re-accretion of the feedback ejecta at coronal and galactic scales (see e.g. \citealt{Semenov2017}). This would then be done at the expense of severe modifications of the galaxies during this timespan, and would then void our approach of controlled experiments. Our restart duration of $30 \myr$ is thus a compromise between not allowing the modifications to induce more side-effects altering the models, and letting the modifications effectively change the star-forming clouds. 

\subsection{Gas density PDF}

\begin{figure}
\centering
\includegraphics{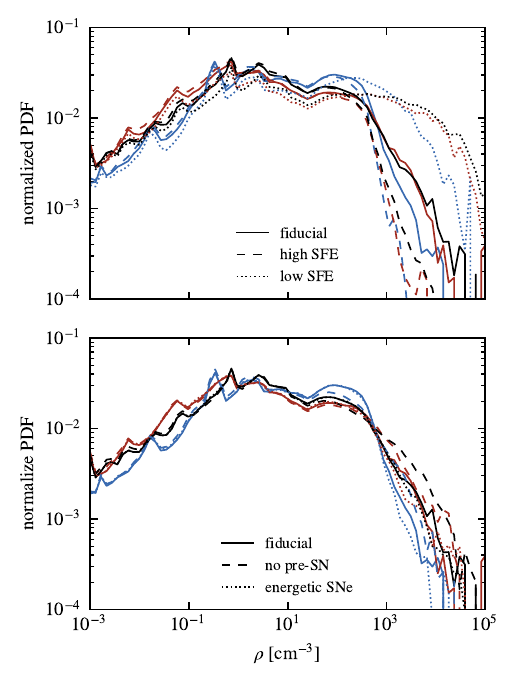}
\caption{Gas density probability distribution function in all our 15 simulations. The color codes the gas fraction (10\% in blue, 25\% in red, 40\% in black), and the line-style shows the modifications from the fiducial cases (solid lines, identical to Figure 3 of \paperi) as indicated in the legends.}
\label{fig:pdf_sfefb}
\end{figure}

\fig{pdf_sfefb} shows the gas density PDFs of the 15 simulations, split for clarity into variations of the SFE (top panel), and of the feedback recipe (bottom panel). For any of the variations considered, gas rich cases (\gm in red and \gl in black) still have slightly larger fractions of dense gas than their \gs equivalent. This difference between high and low \fgas amplifies when reducing the SFE, as the dense gas is consumed at a slower pace. This is however still significantly weaker than the excess of dense gas responsible for the starburst activity in galaxy mergers \citep{renaud2014b,renaud2022, segovia2022}. This increase of the fraction of dense gas is accompanied by a mild deficit of gas at $\sim 0.1\mh 10 \cc$. Separating the atomic and molecular phases as in \paperi reveals that this deficit is found in the atomic outer envelopes of the clumps, which then fuels the molecular centers of the star-forming regions.

Decreasing the SFE from 30\% to 1\% per free-fall time boosts the maximum density by about 2.5 dex. We note that this increase of the maximum density corresponds to a reduction of the free-fall time by a factor of $\approx 20$, which is not enough to fully compensate for the division of the SFE by 30. This indicates limits in the self-regulation of the star formation activity by feedback caused by changes in the ISM density structure.

No strong differences are found when varying the feedback model, neither in the shape nor maximum density. Contrarily to the previous case, this hints towards efficient self-regulation of the star formation process at these time and spatial scales.

In summary, varying the parameters and recipes for star formation and feedback does influence the shape of the gas density PDFs, especially for a low SFE. However, the moderate but significant dependences on the gas fraction found in the fiducial runs (\paperi) are maintained in all cases: the gas-rich galaxies always yield a slightly higher fraction of dense gas than their \gs counterparts, but not a significantly higher maximum density, regardless of the sub-grid recipe used. Gas-rich clumpy galaxies do not contain denser gas, but only a slightly higher fraction of dense gas than contemporary disks, and this trend is largely independent of the star formation efficiency and feedback recipes. 

\subsection{Clump mass function}

\begin{figure}
\centering
\includegraphics{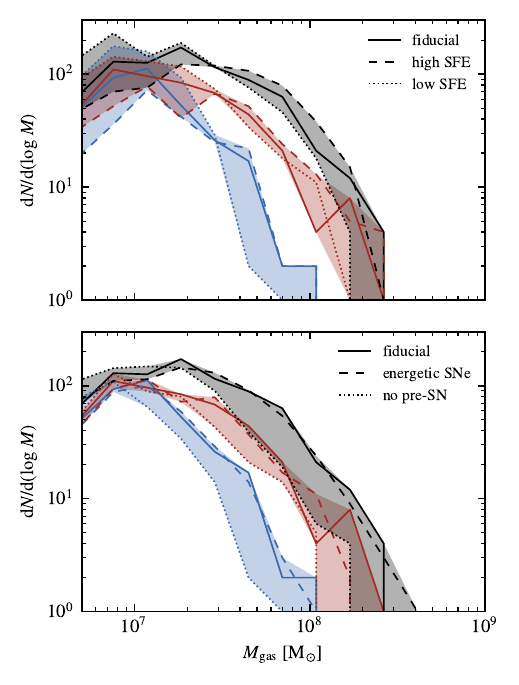}
\caption{Clump mass function, as in \fig{cmf}, in all our 15 simulations. The colors and line styles are as in \fig{pdf_sfefb}. The shaded areas indicate the amplitude of the modifications induced by the change of subgrid models. At the high mass-end, the absence of significant overlap between these shaded areas shows that the gas fraction has a stronger imprint on the distribution of dense clumps than the SFE and the feedback model.}
\label{fig:cmf_sfefb}
\end{figure}

\fig{cmf_sfefb} shows the clump mass function of the 15 simulations, again split into variations of the SFE (top panel), and of the feedback recipe (bottom panel). As for the PDF, our variations in the subgrid recipes, specially with the feedback models, have a weaker influence than the change of the gas fraction. In particular, the models at low-gas fraction remain well-separated than all the others, regardless of the variations of the subgrid models.

\citet{Andersson2024} also varied the subgrid recipes for feedback in simulations of dwarf galaxies. They found that the shape of the \emph{cluster} mass function strongly depends on the timing of stellar feedback, and its effect in stopping the star formation activity in individual clumps. In our analysis, we examine all gas clumps, regardless of their stellar contents. This necessarily implies that our clumps are at different stages of their evolution, including after the onset of star formation. Therefore, the conclusions of \citet{Andersson2024} are not directly translatable in our framework. The mapping of the clump into the cluster mass function is a complex process, subject to both internal alteration from feedback, and external disruptions. As discussed in \sect{disruption}, at least the latter vary with the gas fraction. \sect{profile} complement this analysis by examining the density structure of the clumps, which provides an insight on their resistivity to feedback. Yet, more detailed studies at higher resolution are needed to reach definitive conclusions on the variations with the gas fraction of the transformation of gas clumps into star clusters.

\section{Examples of individual density profiles}
\label{sec:exprofiles}

\begin{figure}
\centering
\includegraphics{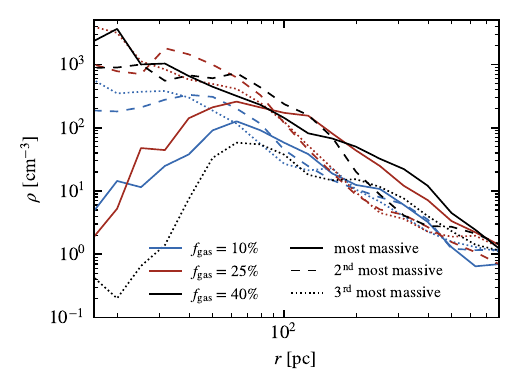}
\caption{Density profiles of the three most massive clumps in each simulation, centered on the center of mass. This illustrates how the diversity of position of density peaks within clumps affects the average profiles shown in \fig{clumpprofile}.}
\label{fig:clumpprofilecom_one}
\end{figure}

\fig{clumpprofilecom_one} shows the density profiles centered on the center of mass for the three most massive clumps of each simulation. The presence of sub-clumps within the same structure shifts the center of mass away from a density peak, such that these density profiles are not monotonically decreasing with radius. When averaging over several clumps, the various radial shifts introduce a statistical flattening of the resulting profile, which is not representative of the structure of any clump. 

However, some of the clumps shown here do yield a flattening of their density profile in its innermost part. The size of this feature is close to our resolution limit ($12 \pc$) and thus subject to caution. Yet, we note that such flat parts extend to different radii from clump to clump, some being captured by up to 6 resolution elements, and do not show rapid variations of the order of the cell size. This suggests that such an inflexion of the density profile could still be detected at higher resolution, and thus be a physical property of massive clumps. More detailed simulations are needed to reach a definitive conclusion.

\section{Correlation between star formation, the clump mass, and the excess velocity dispersion}
\label{sec:masssfr}

\begin{figure*}
\centering
\includegraphics{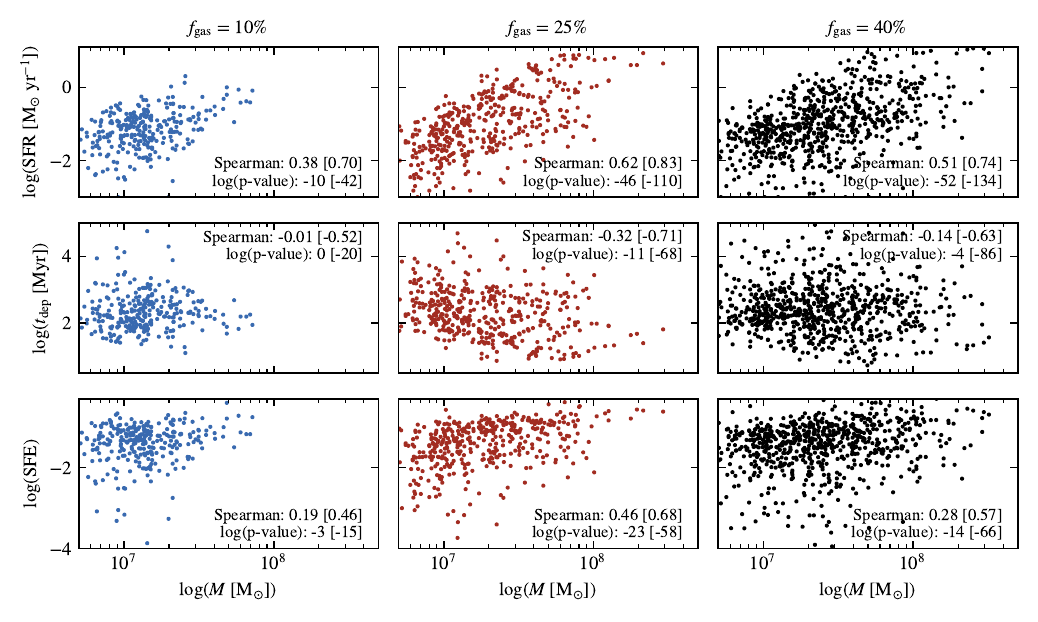}
\caption{Relation between the star formation rate, depletion time, star formation efficiency, and the gas mass of clumps. The number in brackets indicate the correlation coefficient and p-values of the same relations but for the excess mass, copied from \fig{excessmass}.}
\label{fig:masssfr}
\end{figure*}

\fig{masssfr} is the equivalent of \fig{excessmass}, but replacing the excess mass with the mass of the clump. The correlations between the mass and the quantities related to star formation are much weaker (smaller absolute values of the Spearman coefficient) and less significant (higher p-values) than when using the excess mass. Therefore, the mass of gas clumps is a weaker driver of the star formation activity (rate, speed, and efficiency) than the excess of mass with respect to the average clumps of the same size.

\begin{figure*}
\centering
\includegraphics{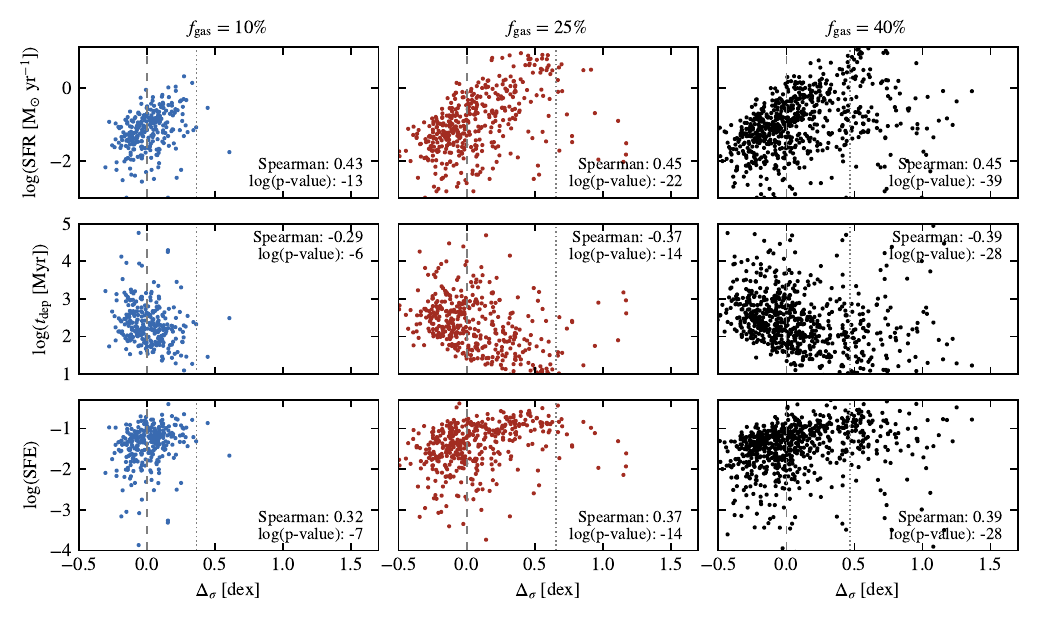}
\caption{Same as \fig{excessmass}, but with the excess velocity dispersion $\Delta_\sigma$ on the horizontal axis. In all panels, the correlations are much weaker and less significant than when considering the excess mass.}
\label{fig:excessvelocity}
\end{figure*}

For completeness, \fig{excessvelocity} show the relation between the indicator of the star formation activity and the excess velocity dispersion. As hinted in \fig{larson}, the correlations are very weak (low Spearman coefficient) and not significant (high p-value).

\end{document}